\documentstyle[12pt,epsfig]{article}
\begin{document}
\begin{center}
{\LARGE Photoproduction of $\eta$-mesons off protons and deuterons}\\[5mm]
C.~Sauermann \footnote[1]
{e-mail:c.sauermann@gsi.de}, B.L.~Friman and W.~N\"orenberg\\[3mm]
GSI, Postfach 11 05 52, D-64220 Darmstadt, Germany\\
and\\
Institut f\"ur Kernphysik, Technische Hochschule Darmstadt,\\
D-64289 Darmstadt, Germany\\[5mm]
\end{center}

\section*{Abstract}
An effective field theory with hadrons and photons is
constructed, where the hadronic parameters are determined by fitting elastic
$\pi$N scattering data. The predicted pion-induced $\eta$-production
on the nucleon
agrees well with the data. The hadronic model is then used to describe
the final-state interaction in photoinduced processes. We present
a consistent description of pion photoproduction in the E$_0^+$-channel
and of the total cross sections for photoinduced
$\eta$-production on the proton as well as on the deuteron.

\newpage

\section{Introduction}

A prerequisite for a theoretical description of the properties of
$\eta$-mesons in hadronic matter and the possible consequences for
$\eta$-production in heavy-ion collisions is a reliable model for
elementary processes like $\eta$-production in hadronic collisions and
in photo-induced reactions.
New accurate data on the photoproduction of $\eta$-mesons off
protons \cite{krusche1,wilhelm} as well as off deuterons \cite{krusche2}
provide strong constraints on models for the elementary
$\eta$-meson--hadron and photon--hadron
interactions.
We present such a model which satisfies
the requirements of unitarity and gauge invariance. An effective
field theory is constructed, where
the hadronic coupling constants and resonance masses
are determined by fitting data on elastic $\pi$N
scattering.
The consistency of our model is checked by comparing the
cross section for the inelastic channel $\pi$N$\rightarrow \eta$N with
experiment (see ref. \cite{paper}).
Having fixed the hadronic parameters of the theory, we now
study electromagnetic processes, namely photoproduction of pions
and etas off nucleons as well as off deuterons.

\section{Photoproduction of $\eta$-mesons off protons}

In the $K$-matrix approach
the $T$-matrix for the
process $\gamma$p$\rightarrow\eta$p  is, to lowest order in the
electromagnetic interaction, given by
\begin{equation}
T_{\eta\gamma}=K_{\eta\gamma} - i \pi \sum_i T_{\eta i} \delta \left(
E-H_i\right)K_{i\gamma},
\end{equation}
where the sum is over all open hadronic channels.
The final-state interaction, included in $T_{\eta i}$, is entirely
due to strong interactions. Under the assumption
that the total cross section for photoproduction of
$\eta$-mesons is dominated by
the S$_{11}$ channel \cite{krusche1},
we use our model \cite{paper} for $\pi$N scattering and pion-induced
$\eta$-meson production in this channel
to describe the final state interaction.
A crucial point is the inclusion of both
resonances in the S$_{11}$ channel, the
S$_{11}$(1535) and the S$_{11}$(1650).
Since both resonances couple to $\pi$N, $\eta$N  and $\pi\pi$N
we include three channels, where the two-pion-continuum
is parametrized by an effective scalar field $\zeta$ (for details
see Ref. \cite{paper}).

In order to satisfy unitarity
we have to include all three channels $i=\pi,\eta,\zeta$ in eq.(1).
We identify the resonance contributions to the $K$-matrix
with the first
diagram shown in fig.~1.
The lagrangian, which describes the coupling of the photon to the
resonance, is given by
\begin{equation}
{\cal L}_{\gamma\mbox{\scriptsize N}\mbox{\scriptsize N}^*}=
\frac{-i e}{2\left(M_R+M_N\right)}\overline{\Psi}_{\mbox{\scriptsize N}^*}
\left(k_R^S+k_R^V \tau_3\right) \gamma^5 \sigma ^{\mu\nu}
\Psi_N F_{\mu\nu} + \mbox{h.c.},
\end{equation}
where $M_R$ and $M_N$ are the masses of resonance and nucleon,
respectively. Since the interaction has an isovector and an isoscalar
part, there are two new coupling constants per resonance.
In order to be consistent with the hadronic part of the model,
which includes interaction terms of the linear sigma model,
and to fulfill the low-energy theorems of pion photoproduction
we also include non-resonant Born terms in $K_{\pi\gamma}$.
The corresponding interaction terms are obtained by coupling the
photon to the lagrangian of the linear sigma model in a minimal
way, which yields a coupling of the photon to the
electromagnetic currents of nucleons and pions, respectively.
Furthermore we add a coupling to the anomalous magnetic
moment of the nucleon and vector-meson exchange contributions.
In order to satisfy the low-energy theorems within the framework
of the linear sigma model in the absence of loop diagrams
an additional
contact interaction of the form \cite{fubini}
\begin{equation}
{\cal L}^c=
\frac{i e g_{\pi\mbox{\scriptsize{NN}}}}{8 M^2}\overline{\Psi}_{N}
\gamma^5 \{\vec{\tau},k^S+k^V\tau_3\}\vec{\pi}\sigma ^{\mu\nu}
\psi_N F_{\mu\nu},
\end{equation}
is added,
which contains an anticommutator of nucleon isospin matrices.
Here $k^S$ and $k^V$ are the isoscalar and isovector part of the
anomalous magnetic moment of the nucleon.
The diagrams shown in fig.~1
are included in  $K_{\pi\gamma}$.
We stress
that the contact term in fig.~1 is not the Kroll-Ruderman term
obtained by coupling the photon in a minimal way
to a pseudovector $\pi$NN-interaction, but corresponds
to the interaction (3). Consequently it is
proportional to the anomalous magnetic moment of the nucleon.
As will be discussed below the vector-meson exchange
processes shown by the last two diagrams of fig.~1 also
play an important role in $K_{\eta\gamma}$.
We introduce form factors for the contact term and the vector-meson
exchange contributions.
Gauge invariance implies large cancellations between the form factor at the
$\gamma$NN-vertex, the one at the $\pi$NN-vertex and nucleon
self-energy insertions \cite{gross,naus}. The net effect is that, for the
current coupling in the Born terms, the form
factors in diagrams 2, 3 and 5 of fig.~1 are cancelled by other contributions.
 \noindent
\setlength{\unitlength}{1mm}
\begin{picture}(150,50)
\put(5,5){\epsfig{file=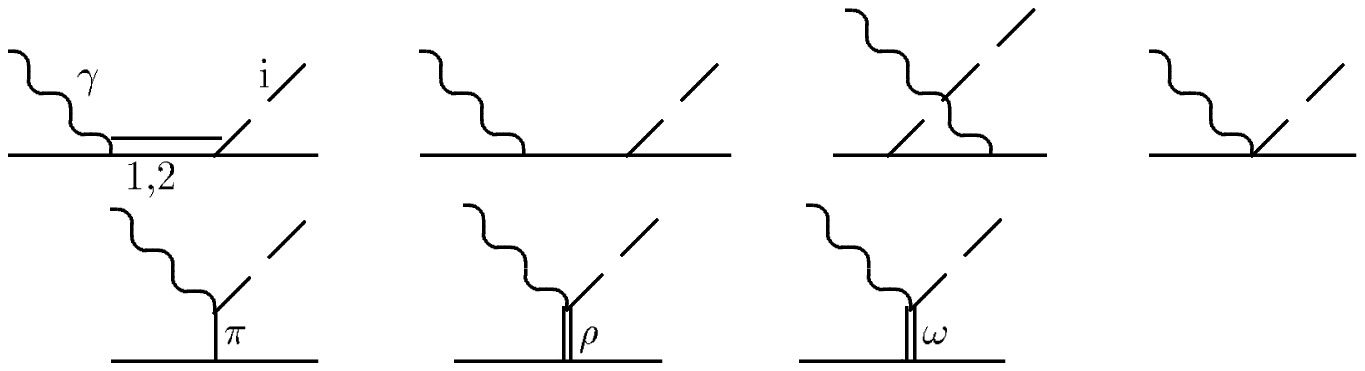,height=40mm}}
\end{picture}

\begin{center}
\parbox{13cm} {\small \it \baselineskip12pt
Fig.~1: Matrix elements $K_{i\gamma}$: The first diagram shows the
contributions from the resonances which contribute
in all three channels (i=$\pi,\eta,\zeta$). The other diagrams correspond
to the $\pi$N Born terms. The last two diagrams contribute to
$K_{\pi\gamma}$ as well as to $K_{\eta\gamma}$.}
\end{center}

Next we have to determine the electromagnetic
parameters, namely the isoscalar and isovector couplings of the photon to the
two resonances and the cutoffs.
The ideal way to proceed would again be to fix the
parameters by fitting the E$_0^+$-amplitude of pion photoproduction
and then make a prediction for the total cross section
for $\gamma$p$\rightarrow\eta$p. However, in view of the fact that
there are substantial deviations between different existing partial
wave analyses for pion photoproduction this is not a practicable
way. Consequently we choose a more pragmatic approach and try
to find a consistent description of the two processes.
Reasonable agreement with
the data on the E$_0^+$-amplitude of pion photoproduction in
the different isospin channels
(not shown) is obtained
for a parameter set which also reproduces the total
cross section for $\eta$-production off proton and deuteron.
We find the following
helicity amplitudes for the two resonances (in units of
$10^{-3}$ GeV$^{-\frac{1}{2}}$):
$A_{1/2}^p$=102, $A_{1/2}^n$=--82 for the
S$_{11}$(1535) and $A_{1/2}^p$=83 and $A_{1/2}^n$=--24 for the
S$_{11}$(1650).
The result for the total cross section for the process
$\gamma$+p$\rightarrow\eta$+p is shown by the solid line in
fig.~2 (left).
The dot-dashed line shows the result if one completely neglects
the presence of the second resonance and the vector-meson exchange
contributions to $K_{\eta\gamma}$ and scales
the result with a factor 0.7 while
the long-dashed line is obtained if one includes the second resonance but
not the vector-meson exchange terms.
Obviously the presence of the
second resonance leads to destructive interference
and a shift of the peak towards higher
energies compared to a model which only includes one resonance.
This shift is compensated by the vector-meson
exchange contributions. We stress that the position of the peak
in the photoproduction cross section is not sensitive to the remaining free
coupling constants, $k_R^S$ and $k_R^V$.
Thus, when the second resonance is included
we cannot describe the data without vector-meson exchange
contributions. This is an important difference to other
models, where
the second resonance was ignored \cite{tiator}.
We note that the coupling
of the S$_{11}$(1650)
to the $\eta$-meson is very weak and consequently
neglected in our model.
In spite of this, the second resonance plays an important role in
the photoproduction of $\eta$-mesons.
In a coupled channel approach this is possible,
in contrast to tree level calculations \cite{mukhopadhyay},
since the S$_{11}$(1650)
couples to the $\eta$N-channel over intermediate $\pi$N, $\zeta$N and
S$_{11}$(1535) states.

\noindent
\begin{minipage}{7cm}
\setlength{\unitlength}{1mm}
\begin{picture}(150,70)
\put(-3,2){\epsfig{file=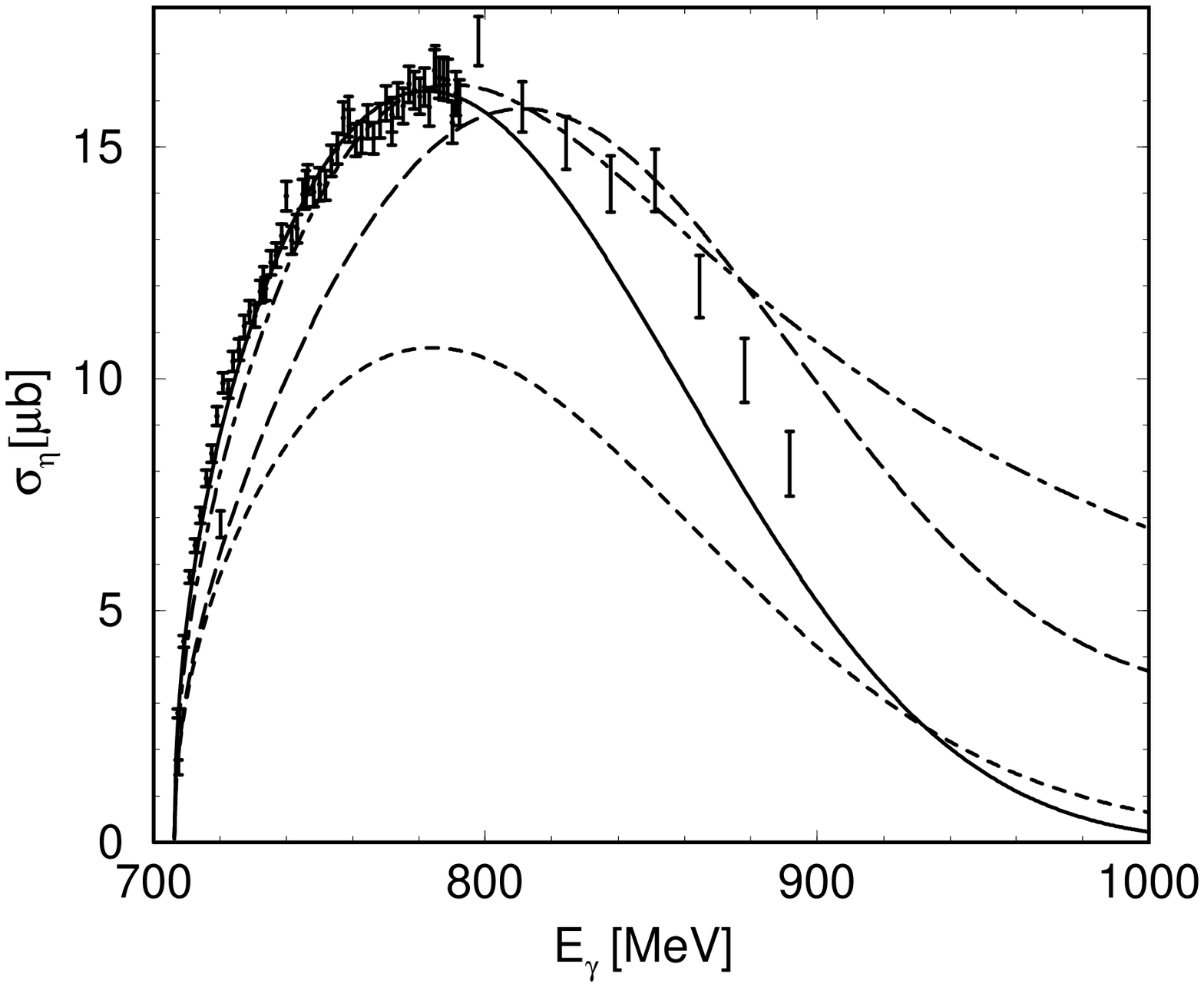,height=70mm}}
\end{picture}
\end{minipage}
\hspace{0.3cm}
\begin{minipage}{7cm}
\setlength{\unitlength}{1mm}
\begin{picture}(150,70)
\put(-3,2){\epsfig{file=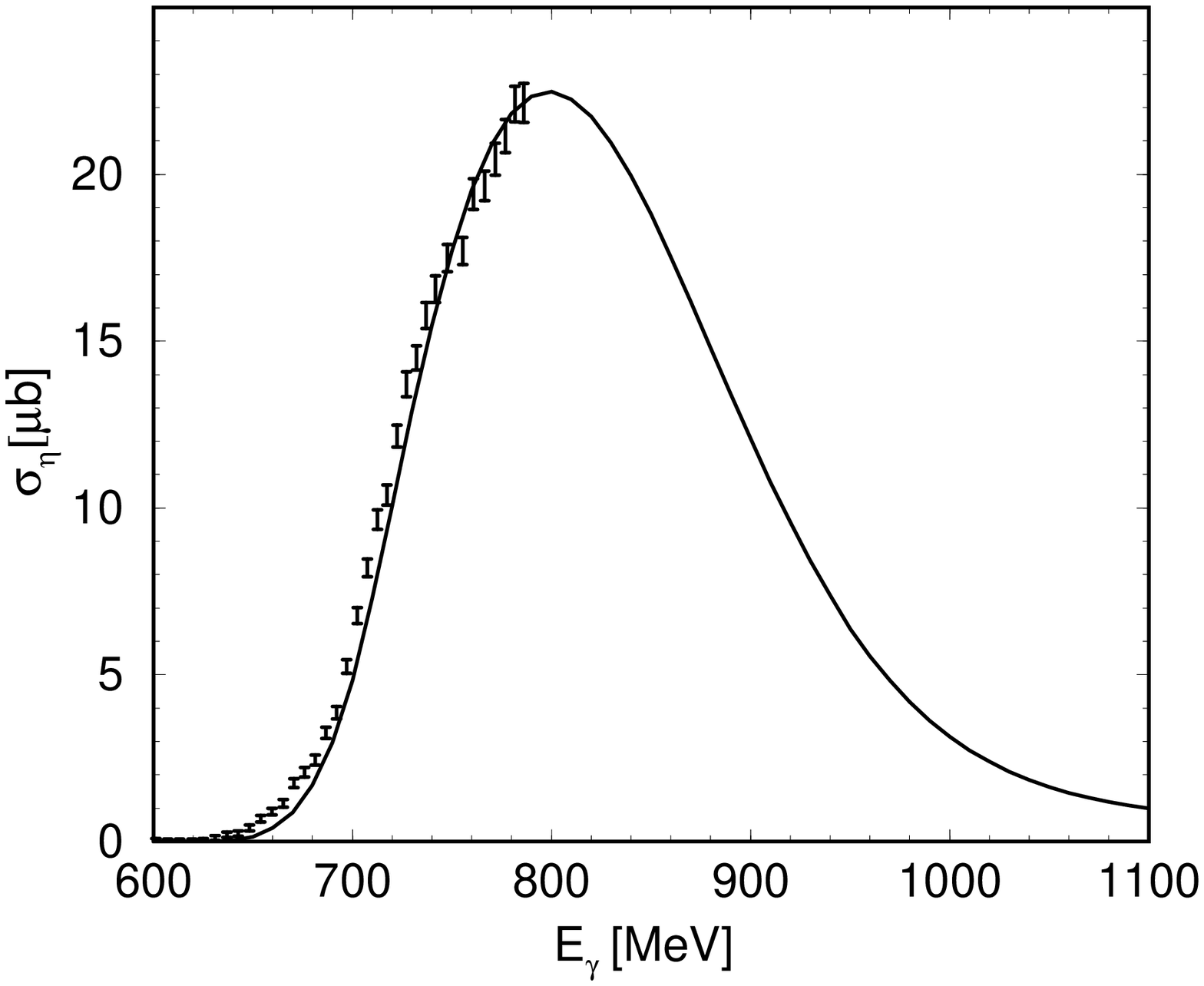,height=70mm}}
\end{picture}
\end{minipage}

\vspace{-0.5cm}
\begin{center}
\parbox{13cm} {\small \it \baselineskip12pt
Fig.~2:
Total cross sections for the photoproduction of $\eta$-mesons
off the proton and off the neutron (left figure, solid and short-dashed line)
and off the deuteron (right figure). The
dash-dotted and the long-dashed curves in the left figure
illustrate the effect of neglecting certain contributions.
For further details see text.
The data are taken from \cite{krusche1,wilhelm} and \cite{krusche2}.}
\end{center}

Also shown in fig.~2 is our
result for photoproduction of $\eta$-mesons
off the deuteron, computed in the impulse approximation.
Except for a deviation near threshold the agreement
with the data is very good.

\section{Summary and Outlook}
The photoproduction of $\eta$-mesons off the nucleon is described
within a unitary and gauge invariant model. Elastic $\pi$N scattering
data are used to determine the hadronic parameters of the model.
The predicted
cross section for pion-induced $\eta$-production agrees well with the data.
We obtain a consistent description of photoproduction of $\eta$-mesons
and pions
satisfying the low-energy theorems for the latter reaction.
We find that the
S$_{11}$(1650)-resonance plays an important role in all the processes
we consider and that the inclusion of $\rho$- and $\omega$-exchange
is crucial in the photoproduction of $\eta$-mesons.

Estimates of the
$\eta$-nucleon pole terms
in the framework of a nonlinear
\linebreak SU(3)$\times$SU(3) sigma model with
mesons and baryons indicate that their contributions
to photoproduction of $\eta$-mesons is small
and can be neglected \cite{weise}. Thus, our model provides
a consistent description of total cross sections
of both pion- and photoinduced $\eta$-meson production. In order to
describe differential cross sections one has to extend the model
by including the P$_{11}$ as well as the D$_{13}$ channel
\cite{tiator, krusche1}.

\end{document}